# DETERMINANTS OF WELL-BEING

Pereira, Cristina1; Gonçalves, Herminia2; Sequeira, Teresa3;
1, University of Trás-os-Montes and Alto Douro (UTAD)
2, 3 Professor and Researcher, University of Trás-os-Montes and Alto Douro (UTAD) & Centre for Transdisciplinary Development Studies

**ABSTRACT**

(1) Background. European social policies have traditionally focused on the material conditions and indicators of well-being and on the foundations of social justice characteristic of the welfare state paradigm. The system of social organisation that ensures a satisfactory standard of living through the provision of social services in areas likely to condition well-being has not sufficiently valued subjective indicators of well-being, making it pertinent to develop international systems of subjective indicators for evaluating well-being.

(2) Objectives and research question. The main objective of this systematic and narrative review of the literature is to substantiate the scientific understanding of the concept of well-being, identifying both the indicators assumed by the literature and by international and national organisations, as well as the relationship with social policy decisions and governance parameters that encourage, affect, and determine well-being. The question this article seeks to answer is this: what can we learn from scientific literature, international guidelines and the cases analysed in the works consulted about the multidimensional relationships established between well-being and governance?

(3) Method. A systematic and narrative literature review was carried out following the PRISMA protocol criteria (search method, screening procedures, document inclusion and exclusion criteria), using the SCOPUS and Web of Science databases, as well as national and international scientific production, articles, research reports, conference proceedings and books, by authors and reference organisations. For bibliometric analysis we used Vosviewer (Rstudio/Bibliometrix software).

(4) Results. The results suggest that: (i) a country's political system, as well as its formal institutions, affect the population's well-being, and the use of well-being indicators in defining public policies is pertinent; (ii) the way political systems are organised can significantly influence citizens' ability to participate in the decision-making process and, consequently, positively affect their well-being; (iii) the use of well-being indicators in politics can fit into a contemporary vision of the role of the state, constituting a promising development that could enable it to fulfil its role in a way that is closer to the interests and real needs of citizens.

(5) Conclusion. This research recognises the importance of using a set of subjective indicators of well-being in addition to income, which are the result of various configurations drawn from the multidimensional relationships that are established between living conditions and well-being.

## 1 INTRODUCTION

The possibility of developing international systems of subjective indicators to assess well-being is on the current scientific agenda. Subjective well-being is understood as the way in which people experience and evaluate their lives overall, or specific domains of their lives (Diener, 1984), while wellbeing refers to a personal state that encompasses good health, adequate material conditions for a safe life, the existence of satisfactory interpersonal relationships and the fulfilment of personal needs that change throughout the life cycle. In this article we will adopt the latter formulation, in which well-being is seen in a more comprehensive way.

Both the delimitation of what well-being means, and its operationalisation have proved to be complex tasks (Dodge, Daly, Huyton & Sanders, 2012). These difficulties are not new. The hierarchy of needs developed by Maslow in the 1950s was probably one of the first theories concerned with understanding how to maximise individual well-being and has been followed by many other attempts to operationalise the concept. As Carrasco-Campos, Moreno and Martínez (2017) rightly point out, sometimes the definition of well-being is associated with the degree of satisfaction of basic needs and the way in which these are met; other times well-being is conceived depending on the country or region analysed. Although there are already tested instruments, the systematic collection of subjective data is still limited, making international comparisons difficult. Understanding and improving citizens' well-being requires a solid evidence base that can inform policymakers about what, when and how to do to improve people's lives. However, to be useful to governments and other decision-makers, data on well-being needs to be collected from large and representative samples, consistently across different populations and groups, over time (OECD, 2013).

In his book "Development as Freedom", Amartya Sen proposes that the notion of development should go beyond considering factors such as Gross Domestic Product growth, income, industrialisation, or technological progress. Sen (1999) argues that development must be associated with an improvement in the lives of individuals and the strengthening of freedoms, pointing to education, health, and civil rights as good examples of catalysing agents for the expansion of real freedoms that all people should enjoy. In this context, the promotion of well-being is, after all, what development is all about, and it must be guided by the search for appropriate answers to a fundamental ethical question: where does the proper value of human life lie? While it is true that there are aspects of any person's life that are valuable in themselves (enjoying good health, acting freely and not being dominated by circumstances, having the opportunity to develop one's potential), there are many "social evils" that deprive people of living minimally well: extreme poverty, collective hunger, malnutrition, social exclusion and marginalisation, deprivation of basic rights, lack of opportunities, oppression and economic, political and social insecurity. For Sen (1999), these variables share the same nature: they are examples of deprivation of freedoms and jeopardise well-being by considering how each person "functions in the broadest sense" (Ai-Thu, 2014).

Since the first cross-cultural studies were carried out, comparing, and analysing variations in the quality of life in different countries (Diener, 1996), it has become clear that well-being is influenced not only by economic factors, but also by other variables such as health or trust in oneself and in institutions. Malinowski's functionalist approach to well-being emphasises the interconnection between culture, social institutions, and the satisfaction of needs. He argues that to understand well-being in society, it is crucial to examine how cultural practices and institutions contribute to social adaptation and stability by influencing human needs. This position has substantiated the opinion of authors such as Bohnke (2006) and Watson, Pichler and Wallace (2009), who have considered that the assessment of quality of life cannot be defined solely through economic and material criteria but must also include the way in which social policies and institutions contribute to the well-being of everyone. Recent research into subjective well-being shows that the development of social and family policies implemented in different countries, aimed at balancing work and family life, help to strengthen the idea of well-being. According to these studies (Moreno, Martínez & Carrasco-Campos, 2016; Segado & López-Peláez, 2013; Watson, Pichler & Wallace, 2009), well-being should be measured using subjective indicators: how individuals feel, how they perceive the notion of well-being, how they assess what is most important in life.

In Portugal, specifically, where there are major inequalities between population groups and territories, an approach of this nature has enormous research potential, particularly in terms of building tools for analysing and monitoring the relationship between people's well-being and social and territorial inequalities.

After presenting the objective, the research question and the methodology followed in this text, in the first section we analyse the extent to which the defence of values such as freedom, participation and trust, as well as the existence of government structures that encourage the civic involvement of their citizens, influence their well-being. In the second section, we look at the concept of well-being from the point of view of the individual, carrying out a critical review of scientific production on well-being in social policies and the indicators commonly used to operationalise it. In the final section we discuss the results and formulate the main concluding ideas.

## 2. OBJECTIVE AND RESEARCH QUESTION

The main aim of this research is to analyse the state of the art on the relationship between social policy decisions and perceptions of well-being. The aim is to establish a scientific understanding of the concept of well-being, identifying both the indicators of well-being assumed by the literature and by international and national organisations, as well as the governance parameters that encourage, affect, and determine well-being. The question this article seeks to answer is therefore: what can we learn from the scientific literature, international guidelines and cases analysed in the works consulted about the multidimensional relationships between well-being and governance?

## 3. METHODOLOGY

Individuals and society are not separable realities, which is why one of the first purposes of research is to generate information that can contribute to a better understanding of social phenomena, starting by identifying the relevant previous research to which this phenomenon relates (Coutinho, 2022; Silva, 2014). The aim of the literature review was to find and localise the most significant studies related to the problem under study. It should be emphasised that the problems investigated are embedded in a social context, reflecting the interests and concerns of the scientific community at a given time.

Given the research objective and question, the methodological option for collecting and analysing data followed the procedures usually used in a systematic and narrative literature review. This review sought to answer the research question posed by constructing a database based on keywords and inclusion criteria, aggregating bibliometric information using the SCOPUS database.

Several authors, including Alves-Mazzotti (2002), consider that narrative literature reviews make it possible to carry out broad analyses and critical interpretations, making it possible to understand the theoretical or contextual point of view of a given subject. Although narrative reviews do not use explicit and systematic criteria for critically analysing the literature, nor do they apply sophisticated and exhaustive collection strategies, they are suitable for understanding the theoretical basis of the problem under study, explaining the legacies of the literature, and allowing the theoretical framework of the research to be expanded.

When consulting the database, the search terms were defined using the keywords "subjective well-being", "social policies", "well-being indicators" and "governance" - in Portuguese and the respective English translations. The selection criteria were documents in the form of articles, books, book chapters, reviews, conference proceedings and research reports from 2014-2022 (Table 1). The database was extracted from SCOPUS for subsequent bibliometric analysis: distribution of publications by year, by country, analysis of citations, co-citations, co-authorship and keyword competition.

In addition, a manual and online search was carried out for national scientific production commonly referred to as "grey literature", namely that produced by authors associated with renowned organisations and/or with recognised qualifications and professional experience. Publications from the last 20 years were favoured in this research, as this was considered an appropriate timeframe for analysing the political transformations in the use of global well-being indicators, since we were looking for the state of the art on the relationship between social policy decisions and perceptions of well-being.

The methods and results of the systematic and narrative literature review should be presented in detail so that users can assess the reliability and applicability of the results of the review.

The Preferred Reporting Items for Systematic Reviews and Meta-Analyses (PRISMA) was developed to facilitate the communication of these reviews and has been updated to PRISMA 2020 to reflect recent advances in the methodology and terminology of systematic reviews, thus facilitating evidence-based decision-making (Page et al., 2021).

We applied the mixed model proposed by Azevedo (2019) to reduce the risk of excluding important scientific references and biasing the results of the review. This choice was made in accordance with the guidelines established by AACODS, developed by Flinders University (Tyndall, 2010), guidelines that assess documents based on criteria such as the reputation of the authors (affiliation to renowned institutions and professional qualifications), methodological accuracy (robustness in the methodology), timeliness of sources (consideration of relevant recent bibliography, as long as it comes from renowned authors), comprehensiveness, objectivity (the literature search focused on welfare terminology, its indicators, political systems and social policies), and significance (the documents were selected for their usefulness and relevance in each of the key words). The systematic narrative review was complemented by the analysis of 48 documents, according to Table 1.

Table 1 - Research guidelines for the systematic narrative review

| Revision | Description | Nº of Documents |
|---|---|---|
| Systematic | 1. Scopus database search using the keywords "subjective well-being", "governance", "social policies" and "well-being indicators" in the title, keywords and abstract | 297 |
| | 2. Chronology defined between | 226 |
| | 3. Selection result: 201 articles, 4 books, 5 conference proceedings | 226 |
| | 4. Eliminated 10 repeated articles and 109 not directly related to the research question | 82 |
| Narrative | 1. Consultation of co-citations and references in the Scopus database | - - |
| | 2. Relevant bibliographical research over the last 20 years | - - |
| | 3. Application of the AACOODS guidelines | - - |
| | 4. Selection of documents: 33 articles, 1 book, 3 conference proceedings, 11 reports from official organisations. | 48 |

The bibliographic data was independently analysed manually by the authors using Vosviewer. The clusters were analysed automatically, considering the distribution of the data, and were complemented by an analysis of the bibliography for each of these clusters.

A narrative review of some cases was also carried out, presenting an analysis of the potential effects of implementing social policies on well-being: at international level, the work of the Stiglitz-Sen-Fitoussi Commission, the OECD Better Life Initiative programme, and the WHO Quality of Life; at national level, the TIWELL - Territories of Inequality and Well-being project.

**4. PRESENTATION OF RESULTS**

4.1.    ANALYSING THE EVOLUTION OF PUBLICATIONS

Between 2014 and 2022, the number of publications about well-being associated with social policies recorded two peak periods, namely in 2019, with 2022 being the second peak period for scientific production in this field (Figure 1).

Figure 1 - Distribution of publications between 2014 – 2022

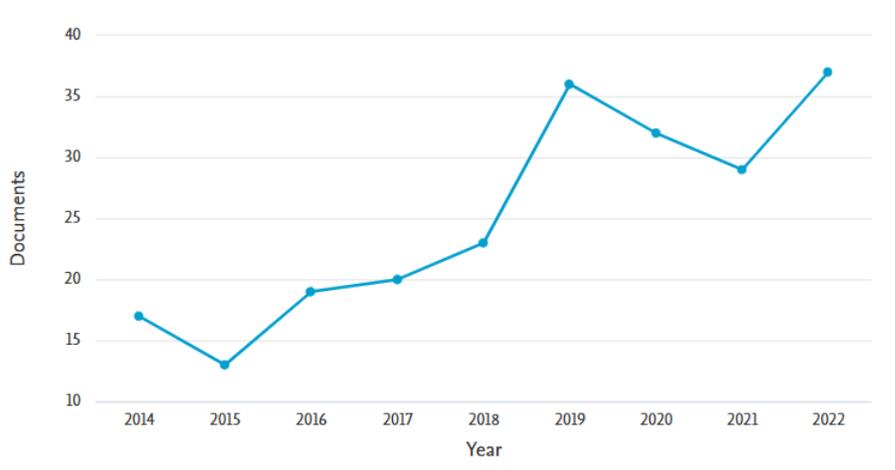

One of the articles, by Orviska & Hudson (2014), called "The Impact of Democracy on Well-being", analysed the impact of governance on well-being using a dataset of world values, from which it emerged that regional democratic satisfaction affects both individual happiness and satisfaction with life, highlighting that democracy has a direct influence on well-being. However, this impact is less significant for women, wealthy people and developed countries.

The analysis revealed that governance varies not only between countries, but also within different regions of the same country. In addition, the research contradicts some previous findings, showing that political freedom contributes to happiness in wealthier countries, while wealthier people can compensate for poor public provision in areas such as health and education, thus emphasising the complexity of the relationship between governance and well-being.

Figure 2 - Countries with the highest number of publications

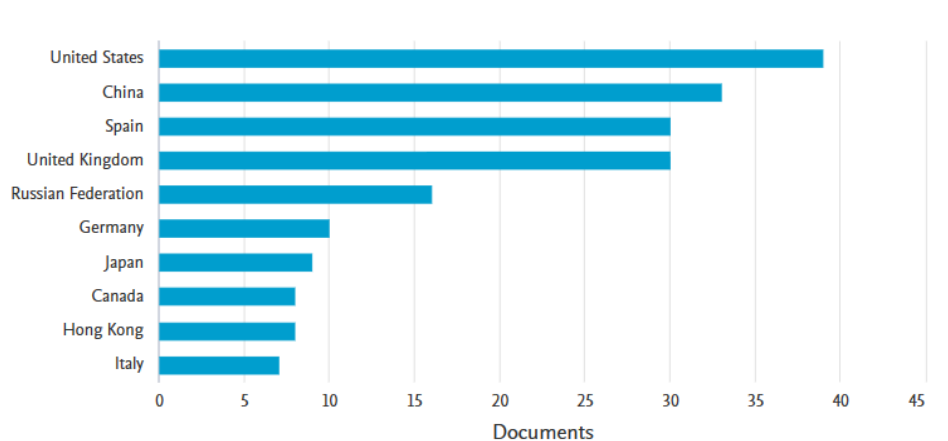

About the 15 countries identified with the highest number of publications, the United States of America stands out (39 documents), followed by China (33 documents) and Spain (30 documents) (Figure 2). Portugal, like Italy, has only one published document.

## 4.2.    CITATION ANALYSIS

The journals show results mainly around Social Sciences (120 documents, equivalent to 29.1%), followed by Psychology (74 documents) and Medicine (44 documents). With the highest number of publications, from the results obtained by Scopus, we can immediately identify the International Journal of Environmental Research And Public Health, with 15 documents, followed by Social Indicators Research (13 documents), Child Indicators Research (7), Journal Of Happiness Studies (5), and Children And Youth Services Review (5).

To analyse the citations per document, a minimum of five citations per article was established. From a total of 201 articles, we selected the five with the highest number of citations (Table 3), having previously listed the five authors with the highest number of publications and citations (Table 2).

Table 2 - Authors with the highest number of publications and citations

| Author | Publications | Citations | Average number of citations per article |
|---|---|---|---|
| Casas, F. | 11 | 179 | 16,27 |
| González-Carrasco, M. | 4 | 49 | 12,25 |
| Abdullah, A.M. | 4 | 50 | 12,5 |
| Gross-Manos, D. | 3 | 70 | 23,33 |
| Montserrat, C. | 3 | 98 | 32,66 |

The most cited article, with 179 citations, presents subjective well-being indicators for large-scale evaluation of cultural services and discusses how these indicators can be used to evaluate the cultural benefits of marine protected areas. Its reflection on the evaluation of cultural benefits as a complex challenge is noteworthy, as it involves the evaluation of subjective experiences and well-being variables, and exploratory factor analysis can be used to identify underlying structures in the data collected on well-being indicators.

The second article suggests a comparison between Seligman's PERMA model and Diener's subjective well-being (SWB) model, seeking to determine whether the PERMA model could identify a unique type of well-being in relation to the SWB model. To provide information on how residents evaluate their own happiness and how these evaluations differ from traditional indicators of well-being, examining the mediating relationships between income and non-income factors, such as quality of life and comparisons of living situations. Investigating the relationship between cultural values and financial and subjective well-being (SWB) through a multidisciplinary approach, the study carried out a meta-analytical review of the field, finding little research focusing on individualism.

These articles address different aspects of subjective well-being, from cultural indicators to comparisons of models and environmental impact, reinforcing the need for multidisciplinary approaches to understanding well-being, integrating cultural, socioeconomic, and environmental variables. They suggest avenues for future research, highlighting gaps in current knowledge about these complex interactions.

Table 3 - The five most cited articles

| Author | Journal | Summary | Number of citations |
|---|---|---|---|
| Bryce, R., Irvine, K.N., Church, A., Ranger, S., Kenter, J.O. | *Ecosystem Services* | Present indicators of subjective well-being for large-scale assessment of cultural ecosystem services and discuss how these indicators can be used to evaluate the cultural benefits of marine protected areas. | 167 |
| Goodman, F.R., Disabato, D.J., Kashdan, T.B., Kauffman, S.B. | *Journal of Positive Psychology* | Comparison between Seligman's PERMA model and Diener's subjective well-being (SWB) model seeking to determine whether the PERMA model could identify a unique type of well-being in relation to the SWB model. | 154 |
| Rivera, M., Croes, R., Lee, S.H. | *Journal of Destination Marketing and Management* | Provide information about how residents assess their own happiness and how these assessments differ from traditional indicators of well-being by examining mediating relationships between income and non-income factors, such as quality of life and comparisons of living situations. | 125 |
| Steel, P., Taras, V., Uggerslev, K., Bosco, F. | *Personality and Social Psychology Review* | Investigate the relationship between cultural values and financial and subjective well-being (SWB) in a multidisciplinary approach. The study conducted a meta-analytic review of the field, found little research, and focused on individualism. | 103 |
| Orru, K., Orru, H., Maasikmets, M., Hendrikson, R., Ainsaar, M. | *Quality of Life Research* | Explore the effect of ambient air pollution on each person's levels of subjective well-being. | 74 |

## 4.3 CLUSTER ANALYSIS

Figure 3 shows the co-citation network, based on the criterion of a minimum of 20 citations per author, resulting in 83 identified authors and 5 clusters. In the red cluster, the most cited author is Diener, E., with 409 citations. The blue cluster with 133 citations is Casas F., followed by the green cluster referring to Oishi S., with 89 citations. The co-citation network shows a strong connection between the clusters, which indicates that the authors tend to cite themselves, despite being part of different disciplinary areas.

About citations per journal, the most prominent are shown in figure 4, after the selection criterion of at least 20 co-citations, resulting in 5 clusters, of which the green cluster for Social Indicators Research stands out, with 2 clusters and 464 citations, as does the same colour for the Journal of Happiness Studies with 219 co-citations and 2 clusters. The blue cluster is followed by the Journal of Personality and Social Psychology with 3 clusters and 147 co-citations.

Figure 3 - Co-citation network by author

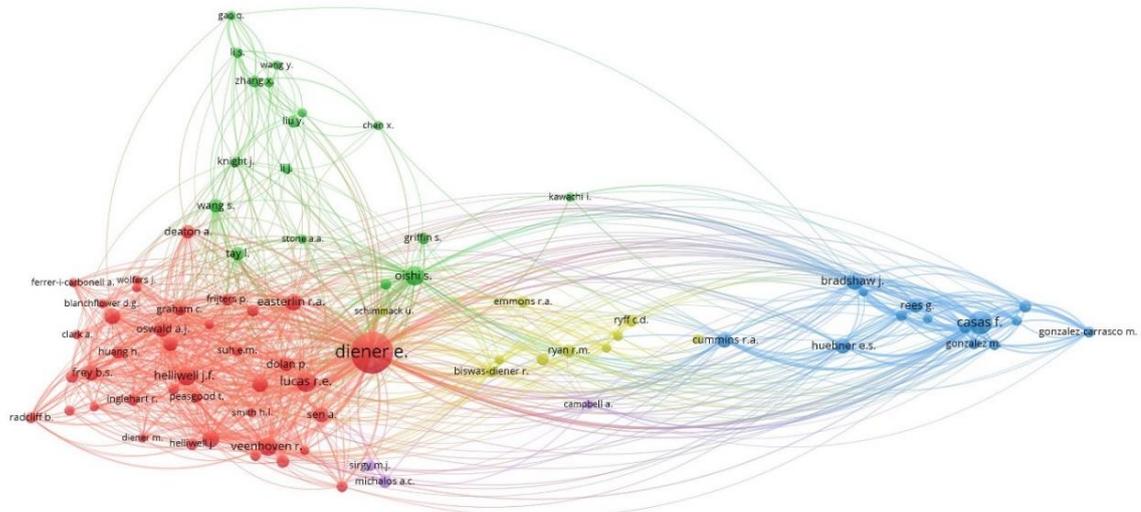

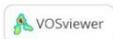

Figure 4 - Co-citation network by journal

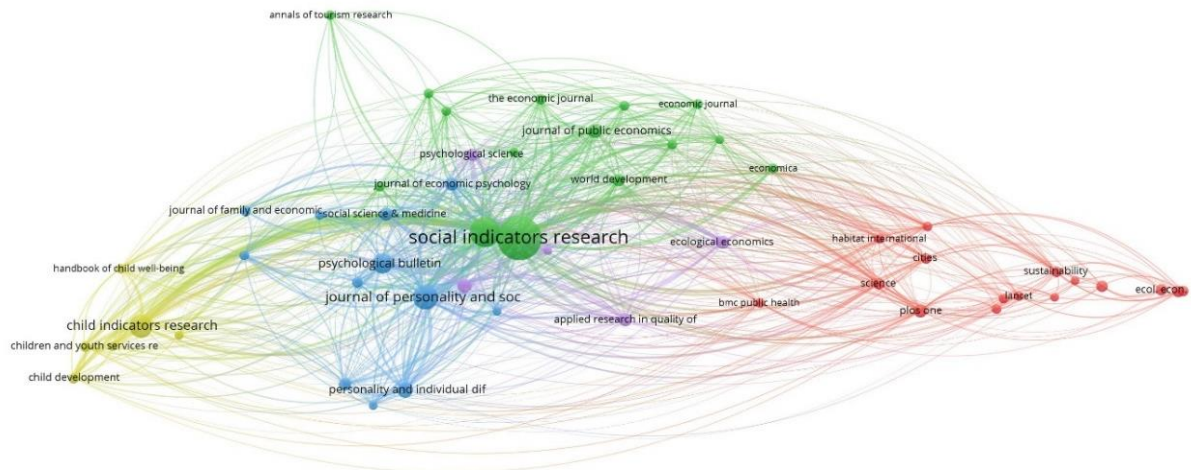

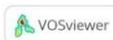

## 4.4 ANALYSING STUDIES ON WELL-BEING AND SOCIAL POLICIES

Below we present a set of studies on well-being (Table 4), carried out in different contexts and by different organisations. Bearing in mind the aim of this study, their analysis is pertinent insofar as they offer us perspectives on the application of well-being measures to political action and illustrate the emergence of the importance of this concept in the field of social policies over the last 20 years.

Table 4 - Studies on well-being

| REPORT | AUTHOR | KEY IDEAS |
|---|---|---|
| *The power and potential of well-being indicators* | Nef – new economics foundation | Research carried out in partnership with Nottingham City Council (UK), which analyzes measures of well-being in the population, particularly personal development. The study demonstrates the importance of this dimension for people's general ability to deal with life's challenges, appearing correlated with physical health, especially at older ages (Marks, 2004). |
| *Happiness, Economics and Public Policy* | Institute of Economic Affairs | It analyzes in detail the economic research that underlies politicians' growing concern with measures of well-being, demonstrating the difficulty in measuring the happiness of societies (Johns & Ormerod, 2007). |
| *Family Database 2013* | Organização para a Cooperação e Desenvolvimento Económico (OCDE) | It evaluates family policies through a wide range of indicators (direct social spending on families, provision of care for dependents, use of parental leave, etc.), through which it seeks to explicitly link family policies with the well-being of recipients of these same policies. |
| *European Social Survey* | ESS ERIC - European Social Survey European Research Infrastructure | European-wide survey (currently covering 30 countries) that incorporates indicators relating to family policies, well-being, and quality of life, and is currently one of the main sources of data on the correlation between well-being and several other social and economic variables. |
| *The social context of well–being* | Royal Society (UK) | It seeks a scientific understanding to create a "science of well-being", resulting in a definition of well-being as a positive and sustainable state that allows individuals, groups, or nations to prosper and grow (Huppert, Baylis & Keverne, 2004); A scientific understanding of well-being is both necessary and timely, making it crucial that governments identify well-being as a more pertinent objective for social development than GDP. |
| *Carta de Genebra para o Bem-Estar* | Organização Mundial de Saúde | It highlights the urgency of creating "wellbeing societies", assuming that wellbeing societies provide the basis for all members of current and future societies to thrive on a healthy planet, regardless of where they live. The Charter also defines that societies must apply bold policies and transformative, sustained approaches, through |

| | | the creation of new indicators of success beyond GDP, that take well-being into account and the definition of new priorities for public spending. |
|---|---|---|
| *The Local Government Act 2000* | Governo de Inglaterra | It marked the introduction of the term "wellbeing" into the concerns of local governments in England and Wales, and it is worth highlighting the development of a Local Government White Paper entitled "Strong and Prosperous Communities", which describes a strategy for political action focused on citizens and where well-being indicators are included in the set of relevant policies to be developed in the future at local level (Steuer, Marks & Thompson, 2007). |

<div align="right">Source: Own elaboration</div>

## 4.5 ANALYSING THE EFFECT OF POLITICAL SYSTEMS ON WELL-BEING

Satisfaction with life is influenced by various social needs and their fulfilment is seen as a prerequisite for happiness and well-being. There has been a growing consensus in the social sciences that income alone is not enough to explain the disparity in well-being and happiness when evaluating social policies. Over the last 60 years, various quantitative metrics have been developed to measure well-being, quality of life and happiness, predominantly using unidimensional scales. Based on a systematic review, the work by Delsignore, Aguilar-Latorre and Oliván-Blázquez (2021) brings together different instruments for measuring happiness and human well-being, with the aim of broadening the literature related to life satisfaction. In particular, the author explores the relationship between tolerance, governance, and inequality in life satisfaction in a sample of 81 countries. While studies have shown that tolerance and governance are separately linked to well-being, no study has demonstrated their mutual relationship with unequal life satisfaction.

The results of the study conducted by Salahodjaev (2021) indicate that more tolerant societies are more likely to have more balanced levels of life satisfaction, but this correlation is completely mediated by the quality of governance. It seems that the quality of institutions is one of the main means by which societies that value tolerance achieve a fairer distribution of happiness. The author also concludes that while GDP per capita contributes to the stability of happiness, income inequality widens the gap in life satisfaction within society. Specifically in this study, the author suggests that tolerance may be indirectly associated with well-being through its impact on the quality of institutions, i.e. governance.

We now analyse the interaction between political systems, the formal institutions of a country, the social policies implemented, and the well-being of the population. We will try to understand what the literature concludes about how a country's institutions (materialised in instances such as government structures, executive bodies, laws, norms of behaviour, conventions, and codes of conduct) interfere with citizens' perceptions of well-being, and the role these formal institutions play in commitment to the political process and strengthening social networks.

In recent decades, the evolution of the institutions that govern the functioning of nations subject to democratic regimes, where freedom is the main virtue, has been marked more by crises of an economic nature than of a political nature. There seems to be a consensus among authors that the inhabitants of freer countries have higher levels of well-being (MacCulloch, 2017; Voukelatou, Gabrielli, Miliou, et al., 2021). Being able to choose is usually considered a positive aspect and, in general, democratic freedoms are seen as vital both for bringing the decisions of governments (national and local) closer to the wishes of citizens and for monitoring and controlling the activity of politicians. Despite its importance, freedom may not be absolute, usually so that the well-being of the majority is not jeopardised by individual desires or religious beliefs. For example, economic freedoms are almost always regulated, namely in the name of increasing social welfare. Free markets should not be confused with unregulated markets, because even so-called liberal markets are governed by a set of laws that guarantee aspects such as property rights, the guarantee of contract enforcement or the quality of the services provided (Acemoglu, 2009).

The study by Serikbayeva and Abdulla (2022) investigated the impact of perceived government performance on well-being, using data from an individual survey in Kazakhstan. This territory, which has undergone significant economic and public administration reforms over nearly three decades since its independence in 1991, offers an interesting setting to examine how people perceive the results of these reforms. This article argues that satisfaction with the quality of public services, living conditions and personal and economic factors influenced by public policies, together with trust in government institutions, play a crucial role in life satisfaction. The more satisfied people are with the quality, accessibility, and cost of public services, the greater their overall satisfaction with life. Furthermore, a high level of trust in government institutions is associated with greater individual well-being.

When revisiting the historical and epistemological foundations of subjective indicators for assessing the well-being of children and adolescents, Casas (2011) highlights the gap in internationally comparable data on the well-being of children and adolescents, suggesting that this can be attributed to the lack of political importance given the perspective of younger people and the lack of consistent research. Despite this, there is rapid progress in research into children and adolescents' views on their living conditions. Although there are tested instruments, systematic data collection is still limited, making international comparisons difficult. Of note is the growing international interest in children's rights to social participation as an opportunity to connect research on child well-being, both objective and subjective. The text concludes by emphasizing the importance of guiding debates on necessary research and challenges to be faced when making research data available to policymakers and public opinion.

Through a cross-sectional analysis of 23 countries from the World Values Survey (WVS), using data relating to the 1980s, Veenhoven (1993) argues that "there is a clear correspondence between the average happiness in nations and the degree to which those nations provide comfort material, social equality, freedom and access to knowledge" (p. 32). Freedom is measured, in this case, by an index relating to political participation, as well as by the existence of a free press. In a later study, Veenhoven (2000) made a comparison between 38 countries based on a single index – combining political, civil and economic freedoms – to again argue that freedom and well-being are positively correlated. However, Inglehart and Klingeman (2000), through a regression analysis applied to well-being variables collected in 105 countries, concluded that "the argument according to which democracy is the basis of well-being is not only in itself sustainable: other factors – particularly (…) the level of economic development of society – appear to play a more powerful role in this domain" (p. 181).

The uncertainty underlying this debate decreases when the influence of the level of economic development, measured by Gross Domestic Product (GDP), on well-being is addressed. In fact, the GDP growth rate has always been considered important (Di Tella, MacCulloch & Oswald, 2003). Inglehart, Foa, Peterson and Welzel (2008), using different waves of the World Values Surveys, carried out between 1981 and 2007, correlated changes in well-being (measured both by perceived happiness and life satisfaction) with the way in which people Political systems and social welfare policies affect this same welfare, having concluded that the evolution of per capita gross domestic product is as important as freedom.

In addition to citizens' disposable income and freedom, the quality of government structures can affect a nation's well-being. Helliwell and Huang (2008) explored variation in life satisfaction data across 75 countries to test the relationship between government quality and well-being, concluding that increasing the overall quality of government functioning would have a similar effect. in life satisfaction, to an increase of around half of the income distribution within the country itself. At stake here are variables such as controlling corruption, encouraging greater citizen participation in public life through the holding of referenda and, in general, the feeling of greater or lesser trust conveyed by government structures.

A growing body of work – namely, the World Happiness Report (Helliwell, Layard, Sachs, et al., 2023), and the OECD Better Life Index (OECD, 2020) – has explored the extent to which certain social responses typical of the State- Provisions, such as the provision of benefits in the event of unemployment or illness, affect well-being. For example, Easterlin (2013, in MacCulloch, 2017) states that if society's objective is to increase people's sense of well-being, then economic growth alone is not enough, and it is necessary to achieve full employment and implement a generous and comprehensive social safety net. For the author, these conditions are indispensable and synonymous with real happiness. In fact, studies show that unemployment situations have a negative impact on well-being, which is even greater than that of a divorce. This is why some economists (Di Tella, MacCulloch & Oswald, 2003) argue that governments should direct all their political efforts towards achieving a situation of full employment or, in its absence, guaranteeing comprehensive social assistance that limits suffering triggered by unemployment, especially when it is prolonged. The existence of generous benefits in situations of unemployment is equally appreciated by both employed and unemployed people, both experiencing an increase of similar magnitude in their respective well-being.

However, doubts persist regarding the overall impact of these benefits on well-being, mainly due to the potential adverse effects on well-being caused by higher taxes (to finance such benefits). Several political scientists question the extent to which the Welfare State provides high levels of well-being given its capacity for people to maintain a "socially acceptable" standard of living, regardless of their participation in the labor market. In 2000, Esping-Andersen (2000) created an index of "decommodification" of work through which he measured the "emancipation of work" in three areas: retirement pensions, income maintenance in case of illness or disability, and unemployment benefits. . A few years later, Pacek and Radcliff (2008) carried out a regression study with data from eleven European nations to test whether or not this "decommodification" index would be correlated with well-being, concluding that, effectively, " "decommodification" of work is positively correlated with life satisfaction. The authors conclude their study with a peremptory statement: "the Welfare State contributes to human well-being".

If it is true that these types of studies seem to support a positive association between the existence of a social safety net and well-being, Veenhoven (2000) argues that public well-being can limit the action of private organizations, as well as lead the loss of individual freedoms in favor of the collective interest. On the other hand, a Welfare State is an enormous consumer of economic resources due to the complexity of its operating structure. For example, Bjornskov, Dreher and Fischer (2007) emphasize that many citizens look with suspicion on the operating expenses of governments typical of countries where the Welfare State prevails, which can negatively affect their perception of well-being related to formal institutions of the respective countries.

**4.6 CASE ANALYSIS: EFFECTS ON THE WELL-BEING OF THE IMPLEMENTATION OF SOCIAL POLICIES**

In this narrative review we will now proceed with the analysis of a series of cases where some potential effects on well-being of the implementation of social policies are presented and discussed.

For many years, social development was essentially measured by traditional indicators of an economic nature, such as GDP, gross revenue, employment and unemployment, income, poverty, or social exclusion rates. However, recent studies suggest the relevance of including indicators of personal satisfaction in evaluating the implementation of social policies (for a review, see Carrasco-Campos, Moreno & Martínez, 2017). Public policies implemented from this perspective, in each country (or even in each region), would play an essential role in promoting the well-being of its citizens.

Following the global economic crisis triggered by the bankruptcy of some North American banks in 2008, the discussion about the way in which wealth is produced and distributed will certainly have contributed to shaking the trust of many citizens in democratic institutions and in decision-makers. politicians. But this discussion will also have contributed to offering a new perspective on the indicators used as a reference to evaluate the performance of the economy, namely, its contribution to social progress and the well-being of populations.

In France, on the initiative of President Nicolas Sarkozy, a working group was created composed of Joseph Stiglitz, Amartya Sen and Jean-Paul Fitoussi, with the aim of identifying indicators that, in addition to quantitative economic data (such as GDP , inflation and unemployment, for example), could be useful and relevant for evaluating social progress (Stiglitz, Sen & Fitoussi, 2009). It is in this context that indicators such as sustainable development, environmental protection and the "quality of life" of populations are highlighted, inaugurating in a way the consideration of the role of political action in promoting well-being beyond traditional economic metrics.

However, following the same strategic line in terms of how to assess the well-being of populations, the Organization for Economic Cooperation and Development (OECD) created the OECD Better Life Initiative program in 2011 (OECD, 2013; OECD, 2020) , which will later integrate and consolidate the work of the Stiglitz-Sen-Fitoussi Commission, whose results gave rise to a publication that today constitutes a reference in this field (Stiglitz, Fitoussi & Durand, 2018).

A broad set of guidelines on the collection and use of well-being measures were produced within the scope of the Better Life Initiative (https://www.oecdbetterlifeindex.org/pt/), aiming to measure society's progress in various areas of well-being, from income and employment to civic engagement and the environment.

The idea of well-being is a central component of this more general framework, and the existence of some type of association between well-being and indicators traditionally linked to life satisfaction and subjective perception of happiness is the subject of specific analysis by the OECD Better Life Index. This index incorporates different dimensions of well-being: income and wealth, employment and salary, housing, health status, work-life balance, education and skills, social and community life, civic engagement, environmental quality, safety, satisfaction with life.

Using data collected in several European countries (Denmark, Sweden, United Kingdom, France, Finland, Netherlands, Spain, Slovenia, Germany, Ireland, Portugal, Greece and Belgium), it was possible to identify indicators relating to each of the dimensions of well-being analyzed, particularly useful information when one intends to study the influence of social policies related to these indicators on the well-being of European citizens (Carrasco-Campos, Moreno & Martínez, 2017). As described on the OECD website, the OECD Better Life Index focuses on aspects of everyday life that matter to people and that shape their well-being. These indicators are regularly updated to better understand well-being trends and their motivations at different historical moments (Table 2).

Table 5 - Well-being indicators of European citizens (adapted from OECD Better Life Index)

| | |
|---|---|
| Income and wealth | Perception of sufficiency to satisfy personal needs |
| | Disposable income |
| | Employment versus unemployment |
| Employment and salary | Wage |
| | Job security |
| Housing | Housing quality |
| | Satisfaction with place of residence |
| Access to social support systems | Health, protection/social assistance and education |
| | Social policies in unemployment situations |
| | Access to formal social support networks |
| Health condition | Health perception |
| | Availability and accessibility to healthcare |
| Work-life balance | Number of working hours and working hours |
| | Reconciling work – family life |
| | Time available for leisure and personal interests |
| | Time available to provide care (to yourself and others) |
| Education and skills | Educational level |
| | Lifelong learning |
| Social and community life | Social support network (friends, neighborhood) |
| Civic involvement | Social participation |
| Environmental Quality | Satisfaction with the environment (air, water, green spaces) |
| Security | Security perception |
| Life satisfaction | Subjective well-being (happiness) |
| | Source: OECD Better Life Index (adapted) |

Still according to the OECD (OECD, 2020) – in the publication "How's Life? 2020 – Measuring Well-being" – between 2013 and 2018, average levels of life satisfaction increased slightly in the 27 countries that comprise it. However, a considerable part of the population (around 7%) still has very low levels of life satisfaction, and around one in eight people experiences more negative feelings than positive on a typical day. Average life satisfaction is very similar for men and women, but in about half of the countries the percentage of women who report more negative feelings than positive is higher than the percentage of men. Countries with greater social inequalities also tend to experience lower average life satisfaction scores.

In this regard, it is worth highlighting here the pioneering work of adapting the World Health Organization's quality of life survey (WHOQoL), adapted for Portugal by Canavarro and collaborators (2006) and which continues to be, even today, a reference assessment of quality of life based on individual perception, widely used in studies in the health area, in particular.

Fundamentally, what is argued in these new ways of evaluating the functioning of society is that traditional indicators of economic performance are insufficient to assess the well-being of populations, something that can only be achieved through a perspective that incorporates the effects of the economy in real conditions. people's lives, populations' vision of their lives and the differentiating role of territories in those same lives (OECD, 2020; Stiglitz, Fitoussi & Durand, 2018). As Mauritti et al. point out. (2022), although the international agenda in this matter is still a polysemous field of intervention, where somewhat equivalent concepts overlap – quality of life, well-being, social progress, development – it is indisputable that we are facing an important and significant impact on the priorities of economic activity. The priority now is people and their real living conditions, with emphasis placed on the processes and mechanisms that shape opportunities for participation and access to the various domains of life beyond income: housing, health, education, work, security, environment, sustainable development.

In Portugal there are great inequalities between population groups and between territories, an approach of this nature presents enormous research potential, particularly in terms of the construction of instruments for analyzing and monitoring the relationships between population well-being and social inequalities and territorial. The TIWELL - Territories of Inequality and Well-being project was designed precisely with the general objective of measuring and monitoring the effects of social inequalities in contemporary Portuguese society and, more specifically, understanding the relationships between social inequalities and well-being in Portuguese municipalities (Mauritti et al., 2022). Another of the objectives of this project is to propose a multidimensional framework for assessing well-being in Portugal, having as its main guidelines the recommendations of Stiglitz, Fitoussi and Durand (2018), the OECD Better Life Index of the OECD and the assessment of operationalized quality of life by Eurostat. For the authors of the TIWELL project, the use of indicators that express individual perceptions is fundamental, reaching through it the way people evaluate dimensions such as healthy living, balance between professional life and family life, the exercise of citizenship, the feeling of security or trust in institutions.

Aware that the profound inequalities in Portuguese society threaten the universality of fundamental rights, the authors of the study (Mauritti et al., 2022) sought to determine to what extent and with what intensity perceptions and experiences of well-being are influenced by asymmetries in living conditions that mark different configurations of territories (interior/coastal; urban/rural), as well as contributing to a sustained framework for locally based public policies and interventions. A system of indicators was developed to measure well-being conditions in different territories of Portugal and interpret their different configurations through an analysis of data on the multidimensional relationships between living conditions and well-being, using quantitative approaches (statistics institutional) and qualitative (Delphi method - case studies). The first phase, already completed, allowed, through a multivariate and multilevel analysis, to define a system of indicators (objective and subjective) for assessing inequalities and well-being in Portugal.

This is a relevant work, as it makes available to the scientific community and political agents, indicators, specifically related to the Portuguese reality, which complement the international indicators usually used in this matter (European Social Survey, OECD, World Bank, European Union) . The first results (Mauritti et al., 2022) demonstrate that the safety and environmental quality of the place where we live, the possibilities for managing family and work time, access to housing and education, health and Transport, along with involvement in communities, are some of the factors that most influence people's perception and experience of well-being, with great variations between territories.

## 6. DISCUSSION OF RESULTS

Here we return to the central question that guides this article – what can we learn, from the data collected and previously presented, regarding the multidimensional relationships that are established between living conditions and well-being – to discuss the potentialities and limitations of using measures of well-being in the definition of public policies.

On the one hand, although sociology and economic science continue to operationalize well-being through material and objective indicators, such as employment, income, housing, health, etc., well-being has become an important concept in the context of European states characterized by adherence to a Welfare State model. Indeed, there is evidence in the literature that demonstrates that access to social policies and economic development can have a significant impact on the well-being of European citizens, but other factors must also be considered when evaluating this well-being, such as interpersonal relationships or the quality of the social and physical environment.

The literature regarding well-being is today extremely extensive and comprehensive, both in terms of proposed theoretical models and in terms of well-being indicators developed to satisfy different purposes (health, environment, etc.). Regardless of the approaches considered, in some cases even competing, Thompson and Marks (2008) argue that the most useful thing for policymakers is to view well-being as a dynamic process, in which circumstances external to the individual interact with their psychological resources to satisfy their needs – to a greater or lesser extent – and give rise to positive feelings of satisfaction and happiness.

On the other hand, the applicability of the concept of well-being in politics and the adoption of well-being indicators as a guideline for political action can be an effective way to improve the well-being of citizens and evaluate the effectiveness of public policies, making It is pertinent to discuss which indicators best assess subjective well-being, particularly in relation to implemented public policies.

Talking about the applicability of the concept of well-being in politics, as well as the adoption of well-being indicators as a guideline for political action, raises some questions that are worth considering. According to some traditions of thought – that is, the liberal tradition – the "happiness" of individuals is not a matter for the state, primarily because there is no "correct" or "universal" model of well-being. But citizens – at least those on the European continent accustomed to living in a welfare state regime – expect the government to promote their interests, and it is possible to argue, in this case, that the use of well-being indicators in politics is consistent with these expectations. This is the point of view of Thompson and Marks (2008), for whom it is also valid to consider that if in some situations we can consider well-being as an end or result of a given policy, in other situations it may be appropriate to think about well-being to achieve desirable or relevant results. Finally, if it is true that most theoretical models of well-being are intended to be applied generally to people of all ages, different indicators of well-being may be differentially more or less important at different points in the life cycle.

A comprehensive look at the set of political initiatives that refer to well-being illustrates what has often been considered a difficulty in using this concept in political action: it's very broad application. For example, well-being is often used as an umbrella term that encompasses a range of positive health behaviors, in this case being perceived as a state of good physical and mental health that can be improved by adhering to certain behaviors. Those who work on economic policy tend to use the term as a synonym for access to various "goods" – economic, health and community resources, political freedom, etc. (Baldock, 2007).

Strongly influenced by so-called positive psychology, the new "science of well-being" aims to determine what factors contribute to a person feeling happy, satisfied, content and fulfilled with their life (Hupert, Keverne & Baylis, 2005). From this perspective, external factors (including physical health and material goods) can play a relevant role in determining the conditions for the emergence and maintenance of a state of well-being but are not equivalent to it. Indeed, the positive psychology research agenda is concerned with trying to identify exactly what variability in well-being can be attributed to factors such as personality, behavior and attitudes, including genetic factors (Lyubomirsky, Sheldon & Schkade, 2005).

For all this, the idea of well-being as a dynamic process is once again highlighted, in the sense proposed by Thompson and Marks (2008). From this perspective, different domains of an individual's life constitute external conditions (work, family, community), which together provide a variety of challenges and opportunities. Psychological resources, in turn, are seen as relatively stable resources and invariant characteristics – personality or self-concept, for example – that influence the way people interact with the external world and respond to its requests. Together, external living conditions and psychological resources support or hinder the satisfaction of needs, leading to an assessment of life in general and/or specific aspects of that life, generating a result corresponding to well-being (Wilkinson, 2007).

But to what extent are these well-being indicators politically acceptable? We know that all measures of well-being depend on individuals' self-reports about aspects of their "inner lives" – thoughts and feelings, emotions, and motivations. In most cases, these aspects are of a strictly individual nature, without any explicit relationship with the functions of central or local governments. Subjective indicators are gaining relevance for policy action, particularly when framing people's views on aspects related to service provision in access deliberations (e.g. 'To what extent are you satisfied with access to healthcare in the local where do you live?') or, on the functioning of society, in terms of perception of safety, for example, in the formulation of products. In these cases, the opinions and feelings expressed may effectively exert some direct influence on those who formulate and legislate public policies. However, it is important to remember that the adoption of well-being indicators as a guideline for political action should not be seen as a simple solution, as it is necessary to take into account cultural, social and political differences in different contexts, and ensure that measures well-being are relevant and meaningful to affected populations.

For example, in Europe we have seen a trend towards greater decentralization in the provision of proximity services. According to Rondinelli (1981), decentralization is the deconcentration of central bureaucratic functions, which presupposes the return and/or transfer, to local governments, of competencies that were the responsibility of the central government or public companies. According to Vásquez, Peñas & Sacchi (2017), in recent decades, many countries have sought decentralization as a means of achieving a more efficient and transparent public sector.

The economic, social, and political consequences of this impulse have been hotly debated, especially to what extent decentralization processes in Europe could contribute to alleviating social inequalities (Greer, 2009). In many states, key competencies related to redistribution and cohesion have been decentralized to the regional level, while macroeconomic and fiscal policy management have been conditioned by European rules. Despite fears that these rules would result in a reduction in well-being and redistribution, and a breakdown in standards of access to public services and well-being, we have actually seen the emergence of regions constructed as new spaces of solidarity and well-being. . The study by Tselios & Rodrigez-Pose (2022) indicates that, in general, decentralization leads to lower levels of national poverty and social exclusion, mainly in European countries with relatively high-quality governance. At the "IV Conference on Public Policies, Planning and Territorial Development – Decentralization and Development", held in Portugal in 2019, several conferences were presented where it was demonstrated that decentralization could effectively reinforce socio-territorial cohesion, allowing to blur the dichotomy between rural and urban (Seixas, 2019), and bring decisions closer to the citizens who will benefit from them (Fernandes & Chamusca, 2019). More recently, Gonçalves and Ferreira (2023) reinforce this idea again, highlighting the role of decentralization in creating new contexts of practice in social intervention at local and municipal level.

Different approaches to measuring well-being can also influence the formulation of public policies. For example, an approach based on economic indicators, such as GDP, can lead to policies that prioritize economic growth to the detriment of other important dimensions of well-being, such as quality of life or the environment. A broader approach considers multiple dimensions of well-being, leading to policies that prioritize quality of life, the environment, and other important dimensions of well-being, in addition to economic growth (Gonçalves, 2022; Keating, 2020).

The relevance of assessing well-being and its influence on the adoption of policies also reaches the essence of an old, but always current, debate about the role of the State in society. According to the influential line of reasoning of political theory commonly described as classical liberalism, the State should not meddle in the personal affairs and choices of individuals, in addition to protecting their private property rights and ensuring that they do not harm anyone with their actions. actions. If people make decisions that harm them in terms of well-being and that make them feel less happy, that is none of the State's business. In Mill's view, a citizen cannot legitimately be compelled to do or not do something because it will be better for him to do so or because it will make him happier (Steuer, Marks & Thompson, 2007). Yet, still, Mill recognized that the functions admitted to a government cover a much wider field than that which can easily be included within the limit of any restriction, justifying in some way that the purpose of governments will be, whenever possible, pursue the interests of citizens.

Knowing that most citizens in Western countries expect their governments to actively promote their interests, this justifies the existence of consistent popular support for the Welfare State model, which is nothing more than a welfare state model. be. The existence of a National Health Service or a universal Social Security regime, in particular, are two good examples of interventions that aim to protect people, and which are clearly based on a concept of promoting and safeguarding well-being as the ultimate end. of government action.

Finally, we learned that different conceptions of well-being proposed in the literature can be useful at different points in the political process, and that different approaches to measuring well-being can influence the formulation of public policies.

## 7. CONCLUSION

Interest in the well-being of citizens has seen a clear increase in recent decades in Europe. Numerous studies have been published highlighting the importance of non-economic indicators related to social policies for the assessment of well-being, focusing on indicators such as compatibility between work and family, personal satisfaction with work, impact of the environment on quality of life, among other variables.

In this work we sought to analyse, through a systematic review of the literature, which multidimensional relationships are established between well-being and governance. Below we present the main conclusions that are important to remember.

Firstly, it became clear that a country's political system, as well as its formal institutions, affect the well-being of the population, and the relevance of using well-being indicators in defining public policies was also highlighted. There seems to be a consensus on the view that a comprehensive social protection network, referred to as a series of policies and programs that aim to protect people against social, economic and health risks, is associated with higher levels of well-being for all citizens.

Secondly, the way political systems are organized can significantly influence the ability of citizens to participate in the decision-making process and, consequently, positively affect their well-being. In fact, the way in which political systems are organized is crucial for citizens to effectively participate in the design of policies and in decision-making processes whose results will directly affect their greater or lesser well-being.

Finally, we conclude that the use of well-being indicators in politics can fit into a contemporary vision of the role of the State, constituting a promising development that could enable it to play its role in a way that is closer to the interests and the effective needs of citizens, promoting a more holistic and integrated approach when formulating public policies.

Well-being indicators can help identify areas where public policies need to be adjusted and improved to meet citizens' needs and interests. This can lead to greater transparency and accountability in political decision-making, evaluating them based on their real impact on citizens' well-being. However, some problems persist regarding the use of well-being indicators to formulate and evaluate public policies, which constitutes a good entry point for future research.

Funding: The paper was funded by national funds, through the FCT—Portuguese Foundation for Science and Technology under the project UIDB/04011/2020.
Institutional Review Board Statement: Not applicable.
Informed Consent Statement:
Data Availability Statement: Not applicable.
Acknowledgments: The authors would like to thank the anonymous reviewers for any review recommendations.
to improve the ideas of the article.
Conflicts of Interest: The authors declare no conflict of interest.

**BIBLIOGRAPHIC REFERENCES**